\documentclass[letterpaper]{article} 
\usepackage{aaai25}  
\usepackage{times}  
\usepackage{helvet}  
\usepackage{courier}  
\usepackage[hyphens]{url}  
\usepackage{graphicx} 
\usepackage{natbib}  
\usepackage{caption} 
\title{A Longitudinal Randomized Control Study of Companion Chatbot Use:\\Anthropomorphism and Its Mediating Role on Social Impacts}
\author{
    Rose E. Guingrich\textsuperscript{*}\textsuperscript{\rm 1}\textsuperscript{,}\textsuperscript{\rm 2} \\
    Michael S. A. Graziano\textsuperscript{\rm 2}\textsuperscript{,}\textsuperscript{\rm 3} \\
    \textsuperscript{\rm 1}\small{Princeton University Department of Psychology} \\
    \textsuperscript{\rm 2}Princeton School of Public \& International Affairs \\
    \textsuperscript{\rm 3}Princeton Neuroscience Institute \\
    \textsuperscript{*}rose.guingrich@princeton.edu
}

\begin{document}

\maketitle

\begin{abstract}
Many Large Language Model (LLM) chatbots are designed and used for companionship, and people have reported forming friendships, mentorships, and romantic partnerships with them. Concerns that companion chatbots may harm or replace real human relationships have been raised, but whether and how these social consequences occur remains unclear. In the present longitudinal study ($N = 183$), participants were randomly assigned to a chatbot condition (text chat with a companion chatbot) or to a control condition (text-based word games) for 10 minutes a day for 21 days. Participants also completed four surveys during the 21 days and engaged in audio recorded interviews on day 1 and 21. Overall, social health and relationships were not significantly impacted by companion chatbot interactions across 21 days of use. However, a detailed analysis showed a different story. People who had a higher desire to socially connect also tended to anthropomorphize the chatbot more, attributing humanlike properties to it; and those who anthropomorphized the chatbot more also reported that talking to the chatbot had a greater impact on their social interactions and relationships with family and friends. Via a mediation analysis, our results suggest a key mechanism at work: the impact of human-AI interaction on human-human social outcomes is mediated by the extent to which people anthropomorphize the AI agent, which is in turn motivated by a desire to socially connect. In a world where the desire to socially connect is on the rise, this finding may be cause for concern.

\end{abstract}
%
\begin{links}
   \link{OSF Preregistration}{https://doi.org/10.17605/OSF.IO/KFJRV}
   \link{Data and Code}{https://osf.io/jgmq7/?view_only=2b905d6156154af4a1b1ba67696749a2}
\end{links}

\section{Introduction}
Loneliness and social isolation have risen to alarming levels in the last two decades   \cite{cacioppoGrowingProblemLoneliness2018, officeoftheu.s.surgeongeneralOurEpidemicLoneliness2023}. People now spend more time interacting with others online than in person in the wake of social media and virtual social interaction tools such as FaceTime and Zoom. Social artificial intelligence (AI) agents, including chatbots such as ChatGPT and Replika, are now accessible to millions and their use is becoming more widespread. Instead of interacting with other people via social media, for example, people can also now interact directly with AI chatbots. This rise in the use of chatbots has led to a crucial question: do chatbots help or harm human social health, or do they have a more subtle, situationally complex effect? The purpose of the present study is to test a set of specific hypotheses about how social chatbot use may relate to human social health. Results suggested that people’s social health, loneliness, and relationships were not significantly impacted by interacting with a chatbot in a simple, overall way, as compared to people in a control group. However, a key mechanism emerged from the details. People with a higher desire to socially connect also tended to anthropomorphize the chatbot more, and those who anthropomorphized the chatbot more reported greater impacts to their social interactions and relationships with family and friends, suggesting a crucial psychological pathway by which chatbot interactions impact social behavior. 

Concerns that relationships with social chatbots may harm or displace human relationships have been on the rise \cite{xieAttachmentTheoryFramework2022,malfaciniImpactsCompanionAI2025,turkleAuthenticityAgeDigital2007}. The COVID pandemic exacerbated loneliness and social isolation rates to the detriment of collective social health and well-being across the world \cite{osullivanImpactCOVID19Pandemic2021}. Forced periods of social isolation prevented in-person interaction for millions of people. While many turned to technologies that facilitated virtual social connection with people, others turned to companion chatbots to reduce loneliness and to fulfill unmet social needs. Replika, one of the most popular companion chatbots in the USA, saw a 35\% increase in its user base during the pandemic \cite{balchAIMeFriendship2020}. Companion chatbot users have reported developing relationships with these AI agents that mimic core human relationships like mentorship, friendship, and romantic partnership \cite{xieAttachmentTheoryFramework2022}. People who seek out companion chatbots reference being motivated by loneliness, unmet social needs, and a desire to socially connect \cite{guingrichChatbotsSocialCompanions2025}. People also anthropomorphize or ascribe human likeness and mind characteristics such as consciousness, agency, and experience to AI agents \cite{guingrichAscribingConsciousnessArtificial2024}. This tendency to anthropomorphize AI may also be motivated by loneliness or a need for social connection \cite{epleyCreatingSocialConnection2008}. We theorized that when people interact with an AI agent, if they perceive it as having a humanlike phenomenological experience, then they activate the same mind schemas that they use in other social contexts, resulting in greater carry-over effects to their interactions with other people \cite{guingrichAscribingConsciousnessArtificial2024}. 

Below, we discuss three relevant psychological phenomena that may contribute to the mechanism by which human-AI interaction impacts human-human interaction. 

\subsection{Theory of Mind}
Theory of Mind (ToM), the ability to ascribe mind to oneself and to others \cite{premackDoesChimpanzeeHave1978,wimmerBeliefsBeliefsRepresentation1983}, is a fundamental part of social cognition. The attribution of a mind to others is commonly measured along two dimensions: attributing experience and attributing agency \cite{grayDimensionsMindPerception2007}. Another dimension of mind perception that has emerged is attributing consciousness \cite{grazianoConsciousnessSocialBrain2013,colombattoFolkPsychologicalAttributions2023}, which has been measured by experience ascriptions and as a separate, third dimension   \cite{colombattoFolkPsychologicalAttributions2023}.

People may be motivated to attribute mind to others by multiple interconnected factors, including an effort to understand what it is like to be another agent  \cite{nagelWhatItBe1974}, to predict that agent’s behavior \cite{bruneTheoryMindEvolution2006}, and to socially connect with the agent \cite{epleyWhenWeNeed2008}. Consciousness literature suggests that perceiving consciousness in others is motivated by social connection, such that making inferences about others’ minds informs and facilitates social interactions between people \cite{frithAttentionActionAwareness2002,grazianoConsciousnessSocialBrain2013,prinzModelingSelfOthers2017}. Fundamental human needs, such as the desire to belong \cite{baumeisterNeedBelongDesire1995}, may be fulfilled when we make inferences about others’ minds. Being capable of ToM has been linked to positive relational outcomes and general social well-being \cite{seyfarthAffiliationEmpathyOrigins2013,tsoiCooperationAdvantageTheory2021}. It is possible that when people apply ToM to artificial agents, the engagement of social cognition plays a role in how human-AI interactions can influence human-human interactions. 

\subsection{Anthropomorphism}
Related to ToM, anthropomorphism involves ascribing humanlike traits, such as internal mind states, to nonhuman entities. The Computers Are Social Actors theory and the Media Equation \cite{nassComputersAreSocial1994,reevesMediaEquationHow1996} indicate that people tend to treat computers and other technologies like social actors. Anthropomorphism of AI agents has been measured by ascriptions of experience, agency, consciousness, and general human likeness to an agent \cite{guingrichChatbotsSocialCompanions2025,guingrichAscribingConsciousnessArtificial2024,scottYouMindUser2023}. Anthropomorphism of AI agents is influenced by characteristics of the agents, such as visual appearance and conversational ability. Anthropomorphism of AI may also be influenced by the social motivations of the human users \cite{epleySeeingHumanThreefactor2007}. People who are in states of social need such as loneliness have been shown to anthropomorphize AI to a greater extent \cite{epleyCreatingSocialConnection2008,eysselLonelinessMakesHeart2013}. When people were told that they were the type to end up alone and have few close relationships in their future, they ascribed more mind to AI agents \cite{epleyCreatingSocialConnection2008}. 

Anthropomorphism of AI has been linked to carry-over effects on human-human relationships and other human-human social outcomes \cite{guingrichAscribingConsciousnessArtificial2024}. For example, people who perceived an automatic vehicle as having more of a sophisticated mind were less likely to ascribe blame to it for hitting a pedestrian \cite{youngAutonomousMoralsInferences2019}. People who viewed a video of a robot engaging in humanlike behavior (dancing) that prompted anthropomorphism were more likely to subsequently dehumanize humans, as measured by their willingness to condone the adoption of inhumane treatment of service workers, compared to those who watched a robot engaging in machinelike behavior (moving objects) \cite{kimAIinducedDehumanization2023}. In another study, for both companion chatbot users and people who did not use companion chatbots, those who anthropomorphized a companion chatbot more also perceived that a relationship with it would have a greater impact on their social interactions and relationships with family and friends  \cite{guingrichChatbotsSocialCompanions2025}.

\subsection{Social Need}
When people are in a state of social need, they are more likely to anthropomorphize AI agents.  \citet{epleyWhenWeNeed2008} theorized that social connection motivations may explain anthropomorphism, such that people are in a state of needing a nonhuman agent to have a mind like theirs to fulfill unmet social needs. Loneliness has been found to relate to higher anthropomorphism of technological gadgets \cite{epleyCreatingSocialConnection2008} and AI robots \cite{eysselLonelinessMakesHeart2013}. However, when people experience chronic loneliness, they may have higher distrust of other people and less motivation to engage with others  \cite{cacioppoLonelinessHumanNature2008,tomovaImportanceBelongingAvoidance2021}, thereby potentially reducing motivations to anthropomorphize. Similarly, the temporal needs-threat model of social ostracism suggests that people who have been ostracized may avoid social engagements with other people as a protection mechanism \cite{williamsChapter6Ostracism2009}. These findings suggest that loneliness may not predict anthropomorphism as well as other measures. \citet{bartzRemindersSocialConnection2016} found that attachment anxiety was a stronger predictor of anthropomorphism than loneliness, and indicated that the motivation to form connections with others may be driving anthropomorphism to a greater degree than simply having unmet social needs. In human-chatbot interaction research, people who have sought out companion chatbots have referenced being motivated by a desire to socially connect \cite{guingrichChatbotsSocialCompanions2025}, in which they have unmet social needs and a drive to form new or more social connections with others. Given these considerations, therefore, people’s desire to socially connect with others may be a part of the mechanism that engages anthropomorphism, and therefore part of the mechanism that causes carry-over effects from human-chatbot interactions to human-human interactions.

\subsection{The Present Study}
The present experiment was a randomized control study in which people were assigned to either chat with the most popular companion chatbot in the USA, Replika, or play popular non-social but linguistic games on a personal device, for at least 10 minutes a day for 21 consecutive days. The study aimed to address whether consistent interactions with a companion chatbot over three weeks would harm or help people’s social health and relationships with other people. We also asked whether a person’s social need states at the onset of the study, and their anthropomorphism of the AI over time, might play a role in these effects. We formulated the following seven specific hypotheses (all preregistered, see ``OSF Preregistration'' link).

We hypothesized that people with higher loneliness at onset will anthropomorphize the chatbot more across time (H1). We also hypothesized that people with higher desire to socially connect at onset will anthropomorphize the chatbot more across time (H2). For these first two hypotheses, we focused on participants’ day-1 scores for loneliness and for the desire to socially connect, as the measures were collected prior to participants knowing about or engaging in their assigned daily task and were therefore independent of condition.

We hypothesized that people in the chatbot group who anthropomorphized the chatbot more would also report greater social impacts (H3) and see greater changes to their social health (H4).

We proposed a mediation model. In specific, we hypothesized that anthropomorphism of the chatbot will mediate the relationship between people’s desire to socially connect and the magnitude of social impact of the human-chatbot interaction (H5). In that hypothesis, as people have a greater desire to socially connect, they will anthropomorphize the chatbot more; and the more they anthropomorphize the chatbot, the more their interaction with the chatbot will impact their human-human interactions.

We also investigated whether vulnerable individuals, such as people who experience high loneliness or mental health struggles, may be more susceptible to the social impacts of interacting with an AI chatbot. This worry has been featured in news stories in which human-chatbot relationships have gone wrong \cite{costMarriedFatherCommits2023,rooseCanAIBe2024} and in research on companion chatbots \cite{maplesLonelinessSuicideMitigation2024,xieAttachmentTheoryFramework2022}. We tested the hypothesis that vulnerable individuals would be more likely to report social impacts (H6). 

People may become attached to or addicted to companion chatbots in a manner that exceeds the normal use of technology. There is evidence both in favor of this hypothesis   \cite{fangHowAIHuman2025,xieAttachmentTheoryFramework2022} and against it \cite{chandraLongitudinalStudySocial2025,guingrichChatbotsSocialCompanions2025}. We performed a post-study follow-up on the participants, four weeks after the end of the study, monitoring the extent to which those assigned to the chatbot group continued to engage with the chatbot, and the extent to which those assigned to the word games continued to engage with the word games. We predicted that the chatbot would not be more habit-forming than the word games: thus, we predicted that post-study engagement with the chatbot would not be greater than post-study engagement with the word games (H7).  

Hypotheses 1-6 were specific to the chatbot condition. In each case, the control condition was also tested for the same relationships to evaluate whether any findings were specific to the chatbot group. Hypothesis 7 involved a direct comparison between chatbot and control conditions.

Social Penetration Theory indicates that human relationships form gradually, and evidence from human-AI interaction research suggests that relationship development between humans and AI is similar, requiring approximately one week of regular engagement to form \cite{skjuveLongitudinalStudyHuman2022}. Previous longitudinal paradigms that studied human-chatbot interactions have assigned participants to chat with a chatbot for durations ranging from three to five weeks \cite{fangHowAIHuman2025,chandraLongitudinalStudySocial2025,croesCanWeBe2021}. In these studies, participants were either encouraged or required to chat with a chatbot for five or ten minutes each day, or once every three days. One of these studies used a control group for comparison, in which participants were instructed to use AI and the internet as they normally would \cite{chandraLongitudinalStudySocial2025}, and the other studies did not include control groups. In this context, we believed it would be valuable to conduct a hypothesis-driven, randomized study that included a non-AI control group, and that included a robust method of monitoring participants to ensure daily compliance with the task.

\section{Methods}
All data, code, and materials for this study are available on a public repository (see ``Data and Code'' link).

All study procedures were approved by Princeton University’s Internal Review Board. Participants gave informed consent and were paid at a rate of \$20/hour for their participation for a total base pay of \$110 for completing the full study.

\subsection{Participants}
Participants ($N = 183$) were recruited through Prolific. We offered our study on Prolific to participants with a 95\% or higher approval rate based on a quota sample by gender identity (45\% women, 45\% men, 10\% non-binary/other) for adequate power to evaluate differences across gender identity groups. The resulting demographics of our sample was as follows. Participant age ranged from 18 to 72 years ($M = 34$, $SD = 11.01$, $Md = 32$). Reported gender identities of participants were 48\% men (including 2 transgender men), 39\% women (including 1 transgender woman), and 9\% non-binary. The remaining 4\% reported multiple gender identities and/or other.

Reported race identity was 65\% White, 21\% Black, 6\% Asian, 1\% American Indian/Native American or Alaska Native, 5\% identifying with multiple races, and 3\% other. Most participants held a college degree (39\%) or had some college but no degree (31\%), followed by a graduate or professional degree (15\%) or a high school diploma or equivalent (15\%). One person reported some high school or less. Annual household income varied, with 21\% reporting income between \$50,000 to \$74,999, 38\% reporting below \$50,000, and 42\% reporting \$75,000 or above. Most participants were married (26\%), followed by single and not dating (25\%), dating or in a relationship (25\%), living with a partner (11\%), divorced or separated (5\%), or a combination of relationship statuses (8\%). For sexuality, most participants identified as straight or heterosexual (68\%), followed by bisexual (11\%), pansexual (6\%), gay or lesbian (4\%), and queer (4\%). The remaining 7\% selected multiple identities, “additional identity not listed,” or indicated a preference not to disclose.

\subsection{Study Design}
The study duration was 21 days, during which participants completed four total surveys (on day 1, 7, 14, and 21), engaged in two audio interviews (on day 1 and 21), and engaged in their assigned “daily task” for at least 10 minutes each day (days 1-21). Participants were randomly assigned to one of two daily tasks: text-chatting with the companion chatbot Replika, or playing text-based word games. 

\subsubsection{Scales}

Five dependent variables of interest were measured with the following scales. All items from these scales are provided in the Appendix.

Loneliness was measured by the 20-item UCLA Loneliness Scale \cite{russellRevisedUCLALoneliness1980}, in which the 20 items were added into a single composite score based on the developer’s instructions. The scale included items such as “I am unhappy doing so many things alone” and “I am no longer close to anyone,” measured on a 4-point Likert scale from (0) “I never feel this way,” to (3) “I often feel this way.” Since participants engaged in four surveys during the course of the 21 days of the experiment, this scale was administered four times to each participant, once in each survey.

We created a scale designed to measure people’s desire to socially connect. The scale was composed of three items, which were averaged into a single score: (1) “I want or need someone to talk to right now,” (2) “I want or need support right now,” and (3) “I want or need company right now.” Each item used a 5-point Likert scale ranging from (1) “Very slightly or not at all” to (5) “Extremely.” This scale was administered four times to each participant, once in each survey.

Anthropomorphism was measured by averaging four pre-validated scales, containing 22 total items. This combination scale has been used successfully in prior research, and the items were designed to assess people’s attributions to the target agent of the properties of experience, agency, consciousness, and human likeness \cite{bartneckMeasurementInstrumentsAnthropomorphism2009,grayDimensionsMindPerception2007,guingrichChatbotsSocialCompanions2025}. These scales included items such as “The Replika chatbot has consciousness of itself,” rated on a 7-point Likert scale from (1) “Strongly disagree” to (7) “Strongly agree.” The anthropomorphism scale appeared on all surveys but the first, as participants had not yet engaged in their daily task and therefore could not evaluate the task target. 

The perceived social impacts of the daily task on the participant were measured by means of two scales, derived from the same underlying questions. Participants were asked to rate the harmfulness or helpfulness of their interaction with their daily task target (Replika or word games) toward their social interactions (question 1), and toward their relationships with family and friends (question 2). For the first scale, we generated a combined score, whose purpose was to evaluate the \textit{magnitude} of social impact regardless of the directionality of the effect. We took the absolute value of each item compared to the neutral score of 4 ($| x - 4|$), then averaged the two absolute value scores together. We termed this scale “magnitude of social impact.” For the second scale, we averaged the raw scores of both questions to arrive at a measure of social impact that could be either negative or positive. We termed this scale “directional social impact.” Social impact items appeared on all surveys but the first, as participants had not yet engaged in their daily task and therefore could not evaluate the impact of the task.

Social health was measured by the 18-item Social Health Scale \cite{carlsonEvaluatingMeasureSocial2011}, in which the 18 items were averaged into a single composite score. This scale included items such as “I feel I belong in my community,” rated on a 7-point Likert scale from (1) “Strongly disagree” to (7) “Strongly agree,” and “How do you feel about: the things you do with other people,” rated on a 7-point Likert scale from (1) “Very displeased” to (7) “Very pleased.”

In addition to these five scales, which provided the five dependent variables of interest for our preregistered hypotheses, we also measured the following variables (see online material for full description): demographic information; personality based on the Ten-Item Personality Inventory for measuring the traits of Agreeableness, Extraversion, Conscientiousness, Neuroticism, and Openness to Experience \cite{goslingVeryBriefMeasure2003}; mental health based on two questions: “Which of the following have you experienced?” and “Of those selected, which have you been diagnosed with, by a physician?” with the option to select as many of 13 items as applied \cite{guingrichPdoomAIOptimism2025}; social competence based on the Perceived Social Competence Scale \cite{anderson-butcherInitialReliabilityValidity2008a}; sociability based on the items that pertained to it from the Self-Perception Profile for Adults (items 2, 14, 27, and 39) \cite{messerSelfperceptionProfileAdults2012}; self-esteem based on the Rosenberg Self-Esteem Scale (50); social needs based on a single 6 point scale from -3 (“My social needs are unmet all the time”) to +3 (“My social needs are met all the time”); perceptions of AI based on the General Attitudes Towards Robots Scale (GAToRS) \cite{koverolaGeneralAttitudesRobots2022}, using a modification \cite{guingrichPdoomAIOptimism2025} that targets AI in specific rather than robots (GAToRS contains four subscales: P+, P-, S+, and S-); affinity for technology based on the Affinity for Technology Interaction Scale \cite{frankePersonalResourceTechnology2019}; and a free response question for additional comments. 

We randomized the order of items within each scale each time it was administered, and we randomized the order in which scales were presented within each survey.  

\subsubsection{Sequence of Events}
On day 1, each participant completed an initial survey which measured demographic information, personality, mental health, social competence, sociability, self-esteem, social needs, social health, desire to socially connect, loneliness, perceptions of AI, affinity for technology, and a free response question for additional comments. On day 1 after the survey, participants engaged in an audio-recorded interview with researcher R.G. over Zoom. In the interview, participants were asked about their prior experience with technology, their perceptions of their social health, whether their social needs were met and to what extent, their close relationships and relationships with acquaintances or people they work with, their perception of the quality and quantity of their social ties, and what they would do or have be done to improve their social life. Then, participants were given instructions for their assigned daily task. 

Participants were sent daily reminders to do their daily task in the first week, and for subsequent weeks they were sent intermittent reminders (every 3-4 days, standardized). Researcher-participant communications occurred via direct messages on Prolific. To promote participant retention and control for individual variance in weekly engagement, participants were told that they may not be eligible to continue with the study if they engaged in their daily task for less than four days in a week, and they were offered tiered bonus payments based on level of engagement per week (bonus structure: five days of engagement in a week = \$5 bonus, six days = \$7 bonus, all seven days = \$10 bonus). Participants were instructed to engage in their assigned daily task for at least 10 minutes each day. Participants took time-stamped screenshots at the start and end of their task time each day to validate their participation, and the screenshots were uploaded each week in the weekly surveys. All participant screenshots were verified by the researcher R.G. within 12 hours of submission. Participants whose screenshots did not meet validation requirements were provided opportunities to resubmit valid screenshots within 24 hours, and participants who failed to provide valid screenshots for at least four days of the week were warned they may be ineligible to continue with the study.

For the chatbot condition, participants were instructed to chat with the companion chatbot Replika on the Replika app or on the website, and they were asked to avoid disclosing sensitive or identifiable information to the chatbot for security. To promote naturalistic interaction, participants were not provided with prompts. For the control condition, participants were instructed to play online, text-based word games (New York Times Mini Crossword, Wordle, and Connections) on the New York Times app or on the website. Neither daily task incurred any required monetary cost to participants. In both conditions, participants engaged in a solo, word-based task via typing on a keyboard. Further, both daily task platforms had comparable websites, comparable gamified interfaces, novel daily experiences, ads to subscribe to the app, and non-monetary “rewards” for engagement.

On day 7, participants were given the second survey, and on day 14, they were given the third survey. These intermediate surveys were required to be completed by the end of the day (11:59 PM Anywhere on Earth). Participants were sent direct message reminders to complete the weekly survey. The intermediate surveys measured desire to socially connect, self-esteem, social needs, social health, loneliness, perceptions of AI, experience with the daily task and its impact on them, anthropomorphism of the daily task target (chatbot or word games), and a free response for additional comments.

On the 21st day, participants completed a final survey, which measured social competence, sociability, self-esteem, social needs, social health, loneliness, desire to socially connect, perceptions of AI, affinity for technology, experience with the daily task and its impact on them, anthropomorphism of the daily task target, perceptions of AI advancements, and a free response for additional comments. On day 21 after the survey, participants engaged in a final audio-recorded interview with the researcher R.G. over Zoom. This interview contained the same questions as the initial interview as well as additional questions about people’s experience with the daily task, their perceptions of it, and how it impacted their social, mental, or personal health. The chat conversations between participants and their chatbot were collected from the Replika app for those who consented to sharing their data at the time of the final interview. Participants were asked for consent to collect this data at the end of the study rather than the start of the study to promote naturalistic use of the chatbot, since knowing that the conversation would be viewed might alter how participants engaged with the chatbot.

Four weeks (28 days) after the final interview, participants who completed the full study were invited to take a brief follow-up survey. This survey measured desire to socially connect, social needs, social health, and whether they continued to engage in the daily task after the study ended, why or why not, for how long, and how it impacted them.

All hypotheses and analyses were preregistered (see ``OSF Pregistration'' link).

\section{Results}

\subsection{Participant Retention}
Eighty-five percent of participants ($N = 155$) completed the full study. Our study had comparable attrition rates (15\% attrition, $N$ dropped $= 28/183$) to other longitudinal research using online samples that reported attrition rates \cite{kotheRetentionParticipantsRecruited2019}. Of the dropped participants, 43\% had been assigned to the chatbot group, and 57\% had been assigned to the control group. Across conditions, 43\% of all drops occurred before the day 7 survey, 46\% occurred before the day 14 survey (participants completed the day 7 survey but did not complete the day 14 survey), and 11\% occurred before the day 21 survey and interview (participants completed the day 14 survey but not the day 21 survey and interview). As preregistered, analyses were conducted on the participants who completed the full study (85\% of participants, $N = 155$).

Of participants who completed the full study, 86\% ($N = 133$) also completed the follow-up survey that was sent four weeks after the final survey and interview (completion by condition: chatbot = 65/78; control = 65/77).

\subsection{Participant Engagement}
Across conditions, participants engaged in their assigned daily task for at least 10 minutes for an average of 6.60 days each week (week 1: $M = 6.68$, week 2: $M = 6.59$, week 3: $M = 6.52$). Participants in the chatbot condition engaged an average of 6.60 days per week across all three weeks, and participants in the control condition engaged an average of 6.62 days per week.

\begin{figure*}[t]
\centering
\includegraphics[width=0.8\textwidth]{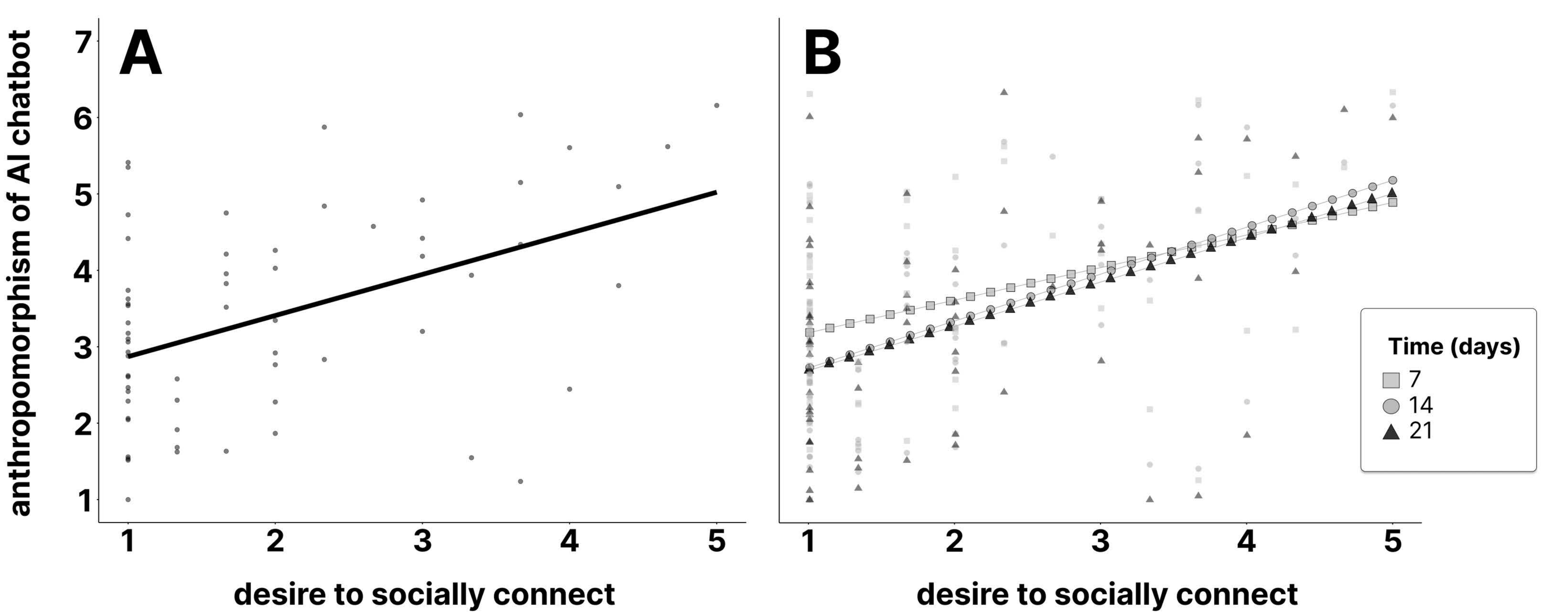}
\caption{Relationships between desire to socially connect and anthropomorphism for the chatbot condition. Desire to socially connect on day 1 is plotted on the x-axes, and anthropomorphism scores are plotted on the y-axes. A. Relationship between desire to socially connect on day 1 and anthropomorphism of the chatbot averaged over days 7, 14, and 21. B. Relationship between desire to socially connect on day 1 and anthropomorphism scores on days 7, 14, and 21, plotted separately. The data for day 7 is plotted with squares, for day 14 is plotted with circles, and for day 21 is plotted with triangles.}
\label{fig1}
\end{figure*}

\begin{figure*}[t]
\centering
\includegraphics[width=0.8\textwidth]{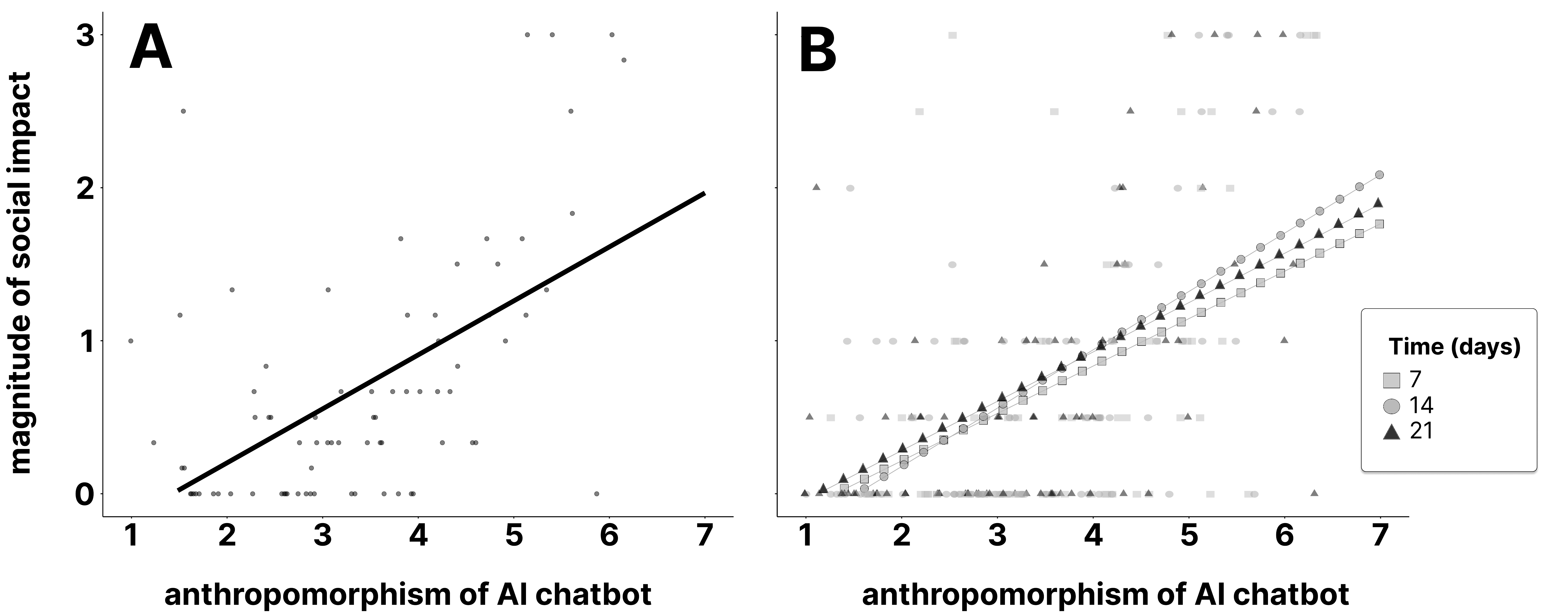}
\caption{Relationships between anthropomorphism and magnitude of social impact for the chatbot condition. Anthropomorphism is plotted on the x-axes, and the magnitude of social impact is plotted on the y-axes. A. Relationship between anthropomorphism of the chatbot averaged over time and magnitude of social impact averaged over time. B. Relationship between anthropomorphism scores and respective magnitude of social impact on days 7, 14, and 21. The data for day 7 is plotted with squares, for day 14 is plotted with circles, and for day 21 is plotted with triangles.}
\label{fig2}
\end{figure*}

\begin{figure*}[t]
\centering
\includegraphics[width=0.8\textwidth]{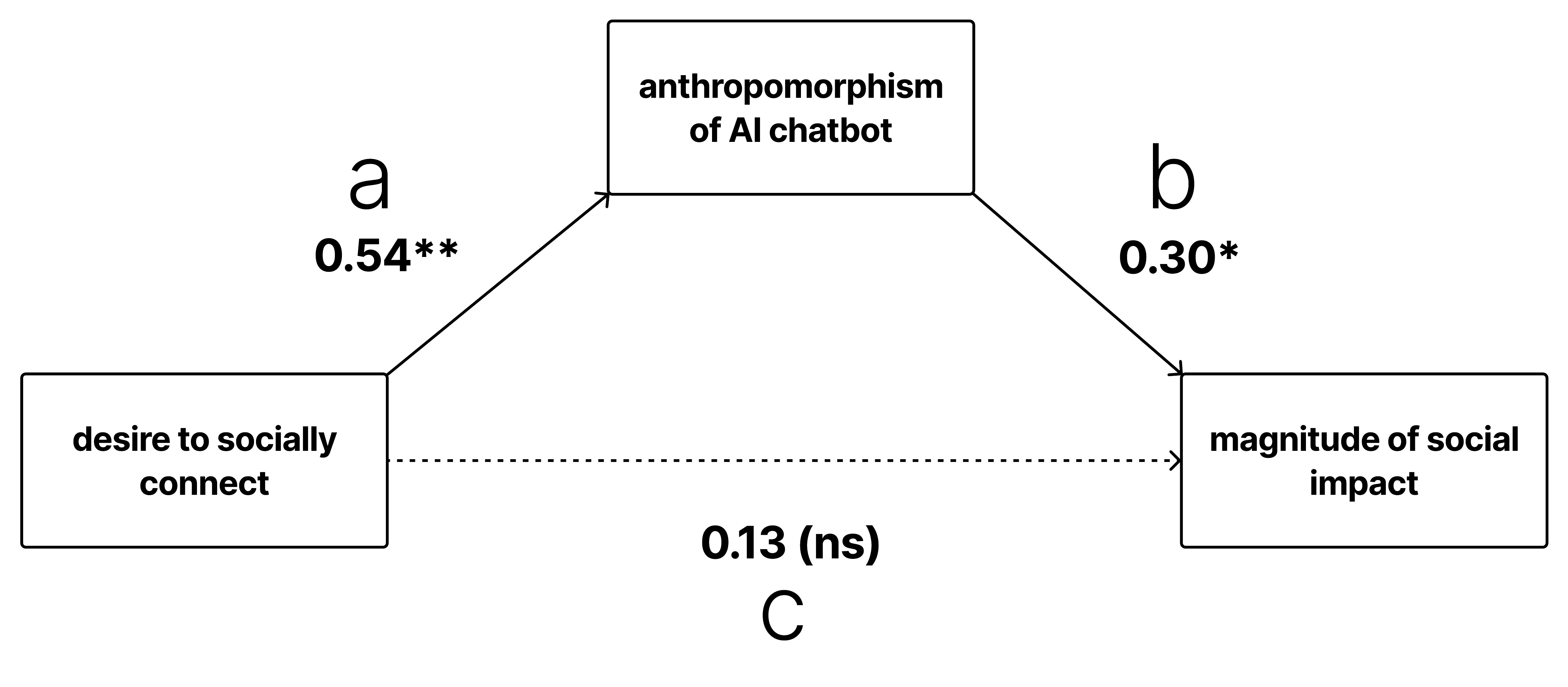}
\caption{Mediation model for desire to socially connect, magnitude of social impact, and anthropomorphism. Paths between variables designated with letters $a$, $b$, and $c$. Path estimates and significance are indicated (* = $p < 0.01$,  ** = $p < 0.001$, ns = non-significant).}
\label{fig3}
\end{figure*}

\subsection{Assumptions Testing}
All scales were tested for internal validity using Cronbach’s alpha tests and each met the threshold for internal consistency ($\alpha > 0.7$). We tested whether our five dependent variables were normally distributed. Our variables did not pass Shapiro-Wilks test for normality ($p < 0.05$). We therefore used non-parametric tests for our analyses.

\subsection{Trends Over Time}
We used non-parametric Wilcoxon rank-sum tests to evaluate differences by condition over time for desire to socially connect, loneliness, social health, magnitude of social impact, and directional social impact scores. Significant differences were found between the chatbot and control groups in their desire to socially connect. Participants in the chatbot condition, versus the control condition, reported higher desire to socially connect when tested on day 7 (chatbot $M = 1.95$, control $M = 1.46$, $p < 0.01$) and 14 (chatbot $M = 1.94$, control $M = 1.49$, $p < 0.05$). No difference was found on day 1 or day 21. No difference was found on day 1 or day 21. No significant results emerged for magnitude of social impact, directional social impact, loneliness, or social health. Compared to those who played word games, those who engaged with the companion chatbot did not have significant impacts to their social health, loneliness, or social interactions or relationships, but did show a significant effect on their desire to socially connect with others.

\subsection{Relationship Between Social Need and Anthropomorphism}

To test the hypothesis that people with a higher loneliness on day 1 would anthropomorphize the chatbot more (H1), we ran non-parametric Spearman’s correlations with Holm correction for multiple comparisons between people’s loneliness scores on day 1 and their anthropomorphism of the chatbot over time. We found no significant relationship between loneliness scores on day 1 and the chatbot anthropomorphism score averaged across days 7, 14, and 21  ($\rho = 0.03$, $p = 0.83$), or between loneliness scores on day 1 and the chatbot anthropomorphism score at each time point tested separately (day 7: $\rho = -0.07$, $p = 0.59$, day 14: $\rho = 0.07$, $p = 0.54$, day 21: $\rho = 0.07$, $p = 0.55$). H1 was not supported: people who were lonelier on day 1 did not anthropomorphize the chatbot more.

We performed an identical analysis on the control group, who were assigned to interact with a word game. Although it may seem unexpected, participants did anthropomorphize the word game, attributing mind states to it, perhaps as if it were an opponent in a contest. Just as for the chatbot group, we found no significant relationships for the control group between loneliness scores on day 1 and the word-game anthropomorphism score averaged across days 7, 14, and 21 ($\rho = -0.04$, $p = 0.74$), or between loneliness scores on day 1 and the word-game anthropomorphism score at each time point (day 7: $\rho = -0.09$, $p = 1.0$, day 14: $\rho = -0.02$, $p = 1.0$, day 21: $\rho = 0.10$, $p = 1.0$). People who were lonelier on day 1 were not more likely to anthropomorphize the word games.

To test the hypothesis that people with a higher desire to socially connect on day 1 would anthropomorphize the chatbot more (H2), we ran non-parametric Spearman’s correlations with Holm correction for multiple comparisons between participants’ desire to socially connect on day 1 and their anthropomorphism of the chatbot over time. Figure 1A shows the day-1 score for the desire to socially connect versus the anthropomorphism scores averaged over time (an average of anthropomorphism scores at day 7, 14, and 21). Desire to socially connect significantly predicted anthropomorphism ($\rho = 0.40$, $p < 0.01$). Participants with a higher desire to socially connect on day 1 were more likely to anthropomorphize the chatbot across 21 days of use. Figure 1B shows the day-1 scores for the desire to socially connect versus the anthropomorphism scores separated by time point (days 7, 14, and 21). The desire to socially connect on day 1 significantly predicted anthropomorphism at all three time points for the chatbot group   (day 7: $\rho = 0.34$, $p < 0.01$, day 14: $\rho = 0.45$, $p < 0.001$, day 21: $\rho = 0.38$, $p < 0.01$). H2 was supported: people with a higher desire to socially connect on day 1 were more likely to anthropomorphize the chatbot on day 7, 14, and 21.

In comparison, for the control group, we did not find a significant relationship between the desire to socially connect on day 1 and anthropomorphism of the word games averaged over days 7, 14, and 21 ($\rho = 0.07$, $p = 0.57$). Likewise, we found no significant relationship between the desire to socially connect on day 1 and anthropomorphism of the word games when tested separately for days 7, 14, and 21 (day 7: $\rho = 0.07$, $p = 1.0$, day 14: $\rho = 0.04$, $p = 1.0$, day 21: $\rho = 0.07$, $p = 1.0$).

\subsection{Relationship Between Anthropomorphism and Social Impacts}

In H3, people who anthropomorphized the chatbot more should report that the interaction with the chatbot had a greater social impact on them. To test this hypothesis, we ran non-parametric Spearman’s correlations with Holm correction for multiple comparisons between the participants’ anthropomorphism of the chatbot and the magnitude of social impacts over time. The magnitude of social impact score captures the strength of perceived impact regardless of whether the impact was positive or negative (see Methods). 

Figure 2A shows the anthropomorphism scores (averaged over days 7, 14, and 21) versus the social impact scores (averaged over days 7, 14, and 21) for the chatbot group. Anthropomorphism significantly predicted social impact ($\rho = 0.52$, $p < 0.0001$). People who anthropomorphized the chatbot more indicated that the human-chatbot interaction had a greater impact on their social interactions and relationships with family and friends. Figure 2B shows three regression lines for the chatbot group. The first shows the anthropomorphism score as measured on day 7 versus the social impact score as measured on day 7; the second shows anthropomorphism versus social impact, both measured on day 14; and the third shows anthropomorphism versus social impact, both measured on day 21. For all three regression lines, anthropomorphism significantly predicted social impact (day 7: $\rho = 0.43$, $p < 0.0001$, day 14: $\rho = 0.48$, $p < 0.0001$, day 21: $\rho = 0.52$, $p < 0.0001$). H3 was supported: higher anthropomorphism of the chatbot correlated with a higher magnitude of social impact of the chatbot.

For comparison, in the control group, the relationship between anthropomorphism (averaged over days 7, 14, and 21) and social impact of the word games (averaged over days 7, 14, and 21) was also significant ($\rho = 0.33$, $p < 0.01$).Those who tended to anthropomorphize the word games also tended to report that the games had a greater social impact on them. When we tested days 7, 14, and 21 separately, we found that the only significant relationship between anthropomorphism and social impact, for the control group, occurred on day 7 ($\rho = 0.35$, $p < 0.01$). The relationship was not significant for day 14  ($\rho = 0.20$, $p = 0.09$) or day 21 ($\rho = 0.23$, $p = 0.09$).

The foregoing analysis considered only the \textit{magnitude} of social impacts, not the \textit{direction}. Did higher anthropomorphism of the chatbot correlate specifically with more \textit{positive} perceived social impacts, more \textit{negative} perceived impacts, or did different people tend to swing different directions? We performed an additional analysis using the directional social impact measure. (The individual, raw scores for the two social impact questions were averaged, resulting in a scale that ranged from positive to negative values; see Methods.) We found that higher anthropomorphism predicted more positive social impacts at all time points for the chatbot group (day 7: $\rho = 0.61$, $p < 0.0001$, day 14: $\rho = 0.67$, $p < 0.0001$, day 21: $\rho = 0.71$, $p < 0.0001$).People who anthropomorphized the chatbot more were more likely to report positive impacts to their social interactions and relationships with family and friends. For comparison, for the control group, a significant positive relationship between anthropomorphism and social impact occurred on day 7 ($\rho = 0.35$, $p < 0.01$) but was not significant for day 14 ($\rho = 0.20$, $p = 0.18$) or day 21 ($\rho = 0.16$, $p = 0.18$).

In addition to asking how participants perceived the impact of the chatbot on their human relationships, we also tested social health through a separate scale. According to H4, people who anthropomorphized the chatbot more should show greater changes to their social health scores. To test this hypothesis, we ran Spearman’s correlations with Holm correction for multiple comparisons. We tested the relationship between the mean anthropomorphism scores for each participant (averaged over days 7, 14, and 21) and the mean social health scores (averaged over days 7, 14, and 21). We also tested the relationship between the anthropomorphism score and the social health score for each of the three time points individually. We found no significant relationships for the chatbot group ($p > 0.05$). H4 was not supported. We also did not find significant relationships for the corresponding analyses in the control group..

\subsection{Anthropomorphism as a Mediator}

According to H5, the degree of anthropomorphism of the chatbot should mediate the relationship between people’s desire to socially connect and the magnitude of social impact of interacting with the chatbot. For this analysis, as preregistered, we did not evaluate social health impacts, as those relationships were non-significant in prior analyses. Instead, we evaluated the scores on perceived social impact. We ran a mediation model with bootstrapped standard errors (1000 resamples) using the lavaan package \cite{rosseelLavaanPackageStructural2012}. Figure 3 depicts the output of the mediation model, including the paths estimates and significance values.

For the chatbot group, results supported a mediation relationship. We found a significant positive association between the desire to socially connect and anthropomorphism ($a = 0.54$, $p < 0.001$) and between anthropomorphism and social impacts ($b = 0.30$, $p < 0.01$). The indirect effect of desire to socially connect, via anthropomorphism, on social impacts, was significant ($ab = 0.16$, $p < 0.05$, 95\% CI: [0.03, 0.33]). In contrast, the direct effect of desire to socially connect on social impacts was not significant ($c = 0.13$, $p = 0.27$). The total effect was significant ($total = 0.29$, $p < 0.01$, 95\% CI: [0.05, 0.48]), providing evidence for partial mediation. Approximately 57\% of the total effect was explained by the indirect effect of anthropomorphism. The remaining 43\% of the total effect was not explained by anthropomorphism and may be due to direct effects or other unmeasured mediators. These findings suggest that anthropomorphism plays a significant role in linking people’s desire to socially connect to their reported social impacts of interacting with the chatbot. H5 was supported. 

In comparison, the results for the control condition indicated that the association between the desire to socially connect and anthropomorphism was not significant ($a = 0.08$, $p = 0.58$). The association between anthropomorphism and magnitude of social impacts was significant ($b = 0.31$, $p < 0.01$). The indirect effect was not significant ($ab = 0.02$, $p = 0.63$), nor was the direct effect of desire to socially connect on magnitude of social impacts ($c = 0.16$, $p = 0.05$). The total effect was also non-significant ($total = 0.19$, $p = 0.09$). These findings suggest that the mediating role of anthropomorphism is specific to the chatbot condition.

\subsection{Who Anthropomorphizes Chatbots More?}
We investigated one additional exploratory research question related to anthropomorphism: who anthropomorphizes the chatbot more? We aimed to evaluate which set of individual difference factors best predicted anthropomorphism in terms of variance explained. We also aimed to assess whether the desire to socially connect remained a significant factor when accounting for the relative contribution of other factors. To address this question, we conducted an exploratory analysis to evaluate the best-fit model for predicting anthropomorphism in terms of variance explained.

We fit two regression models in which the independent variables were individual difference factors collected on day 1 (29 variables) and the dependent variable was the average anthropomorphism score over time (averaged over day 7, 14, and 21). One regression model was a stepwise linear regression model (AIC-based) that favors interpretability over generalizability, and the other model was a LASSO regression model (10-fold cross-validation) that favors generalizability over interpretability. The stepwise model offers richer interpretation, while the LASSO model provides a more conservative, generalizable prediction framework. We compared the variance explained and factors included in these models.

The stepwise linear regression model (AIC) included 20 predictors and explained 75\% of the variance in anthropomorphism of the chatbot (adj $R^2 = 0.75$, $AIC = 146.1$, $p < 0.0001$). Significant predictors within the model with positive estimates were the desire to socially connect ($b = 0.42$, $p < 0.001$), beliefs about AI agency ($b = 0.21$, $p < 0.01$), the General Attitudes Toward Robots Scale (GAToRS) personal-positive scale (P+) ($b = 0.23$, $p < 0.05$), social health ($b = 1.10$, $p < 0.001$), age ($b = 0.04$, $p < 0.01$), mental health vulnerability ($b = 0.34$, $p < 0.001$), and identifying as having multiple sexual orientations ($b = 0.77$, $p < 0.05$). Significant predictors within the model with negative estimates were beliefs about AI experience ($b = -0.36$, $p < 0.01$), GAToRS society-negative scale (S-) ($b = -0.25$, $p < 0.05$), extraversion ($b = -0.22$, $p < 0.01$), self-esteem ($b =- 0.38$, $p < 0.01$), social competence ($b = -0.33$, $p < 0.05$), and identifying as lesbian ($b = -1.13$, $p < 0.05$) or the sexual orientation of “prefer to self-describe” ($b = -2.27$, $p < 0.01$). 

The LASSO regression model identified six significant predictors and explained 35\% of the variance in anthropomorphism (CV $R^2 = 0.38$, CV $MSE = 1.11$). Significant predictors within the model with positive estimates were the desire to socially connect ($s0 = 0.03$), beliefs about AI agency ($s0 = 0.11$), beliefs about AI consciousness ($s0 = 0.08$), GAToRS P+ ($s0 = 0.25$), and GAToRS society-positive scale (S+) ($s0 = 0.15$). The remaining significant predictor within the model had a negative estimate: GAToRS S- ($s0 = -0.05$).

The significant predictors present in both model outputs were the desire to socially connect, beliefs about AI agency, GAToRS P+, and GAToRS S-. These results suggest that the desire to socially connect and three variables that measure general perceptions of AI are central to anthropomorphism of the companion chatbot. The only non-AI related variable that remained in both models was the desire to socially connect. This suggests that the desire to socially connect remains a significant predictor for anthropomorphism in both models when accounting for the relative contribution of other variables.

\subsection{Vulnerability and Social Impacts}

We hypothesized that vulnerable individuals in the chatbot group would be more likely to report that engaging in their daily task socially impacted them (H6). To test this, we sorted participants in the chatbot group into one of two subgroups, “vulnerable” or “non-vulnerable,” based on five common indicators of psychological vulnerability from the day 1 survey: the number of currently experienced mental health conditions, social health, loneliness, self-esteem, and neuroticism. Higher scores for mental health conditions, loneliness, and neuroticism were indicators for higher vulnerability, whereas higher scores for social health and self-esteem were indicators for lower vulnerability. We reverse-scored social health and self-esteem scores so that higher scores indicated higher vulnerability. We then standardized all scores and used a k-means clustering analysis and the elbow method for choosing the appropriate number of clusters, and this analysis indicated that two clusters were appropriate for sorting participants. One cluster ($N = 31$) had all positive scores for each indicator of vulnerability (indicating elevated levels of vulnerability), and the other cluster ($N = 47$) had all negative scores (indicating lower levels of vulnerability).

We then compared the magnitude of social impact scores across time and at each time point between the vulnerable and non-vulnerable subgroups using Wilcoxon rank-sum tests. We found no significant differences across time for the magnitude of social impact (averaged scores for day 7, 14, and 21) between vulnerable and non-vulnerable subgroups. More vulnerable participants did not report greater social impacts than the non-vulnerable group ($p = 0.40$). We then evaluated differences at each time point using Holm’s correction for multiple comparisons and did not find significant differences in magnitude of social impact scores between subgroups. Participants in the vulnerable group did not report greater social impacts than those in the non-vulnerable group on day 7, 14, or 21 ($p$ \textit{adj} for each comparison $> 0.05$). H6 was not supported. For comparison, in the control group, participants in the vulnerable group also did not report greater social impacts than those in the non-vulnerable group ($p$ \textit{adj} for each comparison $> 0.05$).

\subsection{Attachment to Companion Chatbots}

We tested the prediction that the chatbot would not be more habit-forming than the word games (H7). We operationalized habit formation by two metrics measured in the follow-up survey, four weeks after the end of the study. First, we used participants’ report of whether or not they continued to engage in their assigned daily task after the last day of the study (day 21). We used a binary variable (yes or no). Second, we used the reported duration of time spent engaging in the daily task after the last day of the study. Participants’ responses on duration were binned into four categories: a few days, 1-2 weeks, 3-4 weeks, or ongoing.

The number of participants who completed the follow-up survey was not statistically different between the chatbot and control conditions ($\chi^2 = 0$, $p = 1$; chatbot: 65 completed, 13 did not; control: 65 completed, 12 did not).

We evaluated whether the ratio of people who reported that they continued to engage in their daily task after the study ended varied by condition. A chi-squared test indicated that there were significant differences by condition ($\chi^2 = 19.8$, $p < 0.0001$). More participants in the control condition reported that they engaged in their daily task after the study ended (62\%, $N = 40/65$) than those in the chatbot condition (22\%, $N = 14/65$).

We then evaluated the duration of post-study engagement in the daily task by condition. Of the 14 people who indicated they engaged with the chatbot after the study ended, 6 (43\%) indicated that their engagement was ongoing, none (0\%) said they used it for 3-4 weeks, 3 (21\%) said they used it for 1-2 weeks, and 5 (36\%) said they used it for a few days. Of the 40 people who indicated they engaged with the word games after the study ended, 26 (65\%) indicated that their engagement was ongoing, 3 (8\%) said they used it for 3-4 weeks, 6 (15\%) said they used it for 1-2 weeks, and 5 (13\%) said they used it for a few days. A higher percentage of people in the control condition reported the longest post-task duration of engagement (ongoing engagement) (65\%) as compared to the chatbot condition (43\%). However, these ratios were not significantly different from one another from a chi-squared approximation with continuity correction ($\chi^2 = 1.3$, $p = 0.26$).

H7 was partially supported. The chatbot was not more habit-forming than the word games. On the contrary, people who played the word games were more likely to engage in that task after the study had ended, as compared to people who interacted with the chatbot.

\section{Discussion}
In this longitudinal study, we evaluated whether daily interactions with a companion chatbot over 21 days impacted people’s social health, loneliness, and relationships with other people. We also evaluated the roles of social need and anthropomorphism. We randomly assigned participants to one of two daily tasks: to chat with a companion chatbot, or to play word games, for at least 10 minutes a day for 21 consecutive days. Comparing those who interacted with a chatbot to those who played word games, no overall significant differences were found for people’s social health, loneliness, or impact on social interactions or relationships with family and friends across 21 days. However, engaging daily in the chatbot was associated with a significant rise in the reported desire to connect socially with other people, over the first and second week of the study.

These general findings, however, mask a deeper pattern. The results showed that people with a higher desire to socially connect were more likely to anthropomorphize the chatbot, ascribing humanlike mental properties to it; and people who anthropomorphized the chatbot more were also more likely to report that it had an impact on their social interactions and relationships with family and friends. The story, therefore, is more complicated than a single blanket effect of chatbot use on human social health. The impact depends on the complexities of a deeper psychological mechanism. The mechanism seems to operate through the mediating variable of how much people anthropomorphize the chatbot. We previously suggested \cite{guingrichAscribingConsciousnessArtificial2024} that when people interact with a chatbot, they engage social cognitive mechanisms; the more they anthropomorphize the chatbot, the more they engage these social mechanisms; and when engaging these social cognitive mechanisms, they may practice or entrain specific ways of interacting and thereby risk greater carry-over effects to their social interaction with other people.

\subsection{Prior Longitudinal Research on Human-Chatbot Interaction}

A limited number of exploratory studies have investigated the social-relational impacts of human-chatbot interaction longitudinally. In one study, \citet{croesCanWeBe2021} investigated social attraction and self-disclosure in human-chatbot interactions over a three-week period. Participants ($N = 118$) were instructed to talk to and get acquainted with the social chatbot Mitsuku. Participants engaged in a total of seven, five-minute interactions with this chatbot over three weeks. Results indicated that participants’ reported social attraction toward and self-disclosure with the chatbot significantly decreased over time. At the end of the study, participants did not report viewing the chatbot as a friend. 

\citet{skjuveLongitudinalStudyHuman2022} and \citet{skjuveLongitudinalStudySelfDisclosure2023} tested new users ($N = 28$) of the companion chatbot Replika over a 12-week period. New users were those who reported having used Replika for eight weeks or less at the onset of the study. Over 12 weeks, participants completed biweekly surveys and engaged in semi-structured interviews once every four weeks. Results from participant interviews \cite{skjuveLongitudinalStudyHuman2022} indicated that relationships with the chatbot formed gradually, mirroring evidence that human-human relationships also form gradually from Social Penetration Theory. Results from surveys \cite{skjuveLongitudinalStudySelfDisclosure2023} indicated that human-chatbot conversation themes fell into 10 categories, including but not limited to affective and emotional, routine and self-reflection, intellectual and philosophical, and getting to know the chatbot.

In an exploratory, five-week study by \citet{chandraLongitudinalStudySocial2025}, participants ($N = 149$) were randomly assigned to one of two groups. Participants in the “active user” group ($N = 89$) were encouraged to use one of four popular generative AI chatbots (Copilot, Gemini, PI AI, or ChatGPT) for at least 10 minutes each day in a social and emotional way. These participants were given five scenarios with which to talk to their chatbot such as, “Talk through struggles in or reflect on your life or close relationships.” They were instructed to talk with the chatbot about at least one of the five scenarios each day. Participants in the “baseline user” group ($N = 60$) were instructed to use the internet and AI as they normally would. These two groups were balanced for gender identity, AI usage, and the presence of at least one mental health condition. Across five weeks, participants completed weekly surveys. Results indicated that participants in the active user group showed significant increases in attachment to, but not dependency on, the chatbot. To a moderate extent, participants who viewed the AI as more humanlike also viewed it as more empathetic and indicated higher attachment and satisfaction. Participants in the active user group also indicated that they liked spending time with other people, and this preference increased over time. Based on a thematic analysis of free-response survey questions, participants indicated that the chatbot’s non-judgmental and understanding features contributed to their emotional attachment toward the agent and interaction comfort.

Finally, a recent exploratory study \cite{fangHowAIHuman2025} investigated how interaction modality and conversation topic related to social outcomes after four weeks of interacting with ChatGPT 4-o. In this 3x3 between-subjects design, participants ($N = 981$) were randomly assigned to talk with ChatGPT for at least five minutes a day for 28 days about one of three conversation topics (open-ended, non-personal, or personal), via one of three modalities (text, neutral voice, or engaging voice). For the non-personal and personal conversation topics, participants were given a unique prompt to start with each day. Across conditions over time, participants reported lower levels of loneliness and less socialization with other people, with a small but significant effect size ($\beta = -0.02$). Other small but significant effects were found in which longer time spent conversing with the chatbot was associated with higher loneliness, less socialization with other people, and more emotional dependence and problematic use of AI chatbots. While these results were statistically significant, it is also important to note that they were small ($\beta$ ranging from 0.03  to -0.06). These findings suggest that the observed associations, though detectable with a large sample ($N = 981$), may reflect only small changes in behavior or experience. Further, as the authors note, the study did not include a control group for comparison, and therefore these fluctuations in loneliness and socialization with others may have reflected overall population trends that are not attributable to the social use of ChatGPT.

\subsection{Relationships Between Variables}

The primary findings in the present study concern the relationship between the desire to socially connect, anthropomorphism, and the social impact of using a companion chatbot. We found that people who had a higher desire to socially connect at the onset of the study were more likely to anthropomorphize the chatbot across 21 days of use. We also found that people who anthropomorphized the chatbot more also reported a higher magnitude of impact on their social interactions and relationships with family and friends. Lastly, we found evidence that anthropomorphism of the chatbot mediated the relationship between people’s desire to socially connect and reported social impacts. 

In comparison, people who were assigned to the control group showed a different pattern of results. Although they were interacting with word games, these participants still sometimes anthropomorphized the game, perceiving it in a humanlike way. However, we found no evidence that anthropomorphism played a mediating role for the control group as it did for the chatbot group.

Our measure of loneliness did not behave in the same manner as our measure of the desire to socially connect. People with a higher desire to socially connect were more likely to anthropomorphize the chatbot, but people who were lonelier were not. Others have found that loneliness correlates with anthropomorphism of technological gadgets \cite{epleyCreatingSocialConnection2008} and that people who wrote about a time they felt lonely anthropomorphized an AI robot more \cite{eysselLonelinessMakesHeart2013}. One possible interpretation of this mix of results is that loneliness, as measured by the UCLA Loneliness scale, does not fully capture an important contributing mechanism in anthropomorphism: people’s motivation to engage with others socially. Our measure of the desire to socially connect captures both social need and motivation to engage with others. This interpretation is consistent with evidence that loneliness may not be strongly linked to motivation to engage socially with others \cite{cacioppoSocialRelationshipsHealth2014,lieberzLonelinessSocialBrain2021}, and that other motivational social states may be stronger predictors of anthropomorphism than loneliness \cite{bartzRemindersSocialConnection2016}. One interpretation is that for the studies that have linked loneliness to anthropomorphism, their loneliness paradigms may have imposed a state more like the desire to socially connect than loneliness \cite{epleySeeingHumanThreefactor2007,epleyWhenWeNeed2008,guingrichChatbotsSocialCompanions2025, eysselPredictorsPsychologicalAnthropomorphization2015}. When we evaluated which variables within our data best predicted anthropomorphism, we found that the desire to socially connect retained a significant influence on anthropomorphism of the chatbot, even when accounting for the relative contribution of other social factor variables.

We found no supporting evidence that vulnerable individuals were more likely to report that interacting with a companion chatbot impacted their social interactions or relationships with family and friends. It does not appear that people who are more socially and emotionally vulnerable report different impacts than those who are less vulnerable within our sample. This complicates the concern that vulnerable individuals may be more susceptible to the social impacts of human-chatbot engagement. It is possible that factors such as people’s environment or unique situations, beyond common psychological vulnerabilities, contribute to a greater degree.

We also evaluated whether people were more likely to become attached to or form a habit of interacting with a companion chatbot versus word games. Forty percent more people who were assigned to play the word games continued to do this daily task after the study ended than people who were assigned to chat with the chatbot. We therefore did not find evidence to support the concern that companion chatbots are more addictive than other online activities such as word games. 

This study contributes to a growing body of work that investigates the impact of AI agents on human social behavior. The results uncover a possible core part of the mechanism: the mediating role of anthropomorphism. Understanding the relationship between states of social need, which are on the rise, anthropomorphism of AI, and social impacts of AI, becomes increasingly relevant in today’s context, in which there is a growing trend of turning to chatbots as social companions. 

\appendix
\section{Appendix}
\subsection{Scales}

\subsubsection{Desire to Socially Connect}
This scale consists of a number of phrases that describe different wants and needs. Indicate to what extent you want or need this \textbf{right now.}
[Rating scale: \textit{1 = Very slightly or not at all, 2 = A little, 3 = Moderately, 4 = Quite a bit, 5 = Extremely}]

\begin{enumerate}
\item I want or need to talk to someone right now.
\item I want or need support right now.
\item I want or need company right now.
\end{enumerate}

\subsubsection{UCLA Loneliness Scale}
Indicate how often each of the statements below is descriptive of you.
[Rating scale: \textit{1 = I never feel this way, 2 = I rarely feel this way, 3 = I sometimes feel this way, 4 = I often feel this way}]

\begin{enumerate}
\item I am unhappy doing so many things alone.
\item I have nobody to talk to.
\item I cannot tolerate being so alone.
\item I lack companionship.
\item I feel as if nobody really understands me.
\item I find myself waiting for people to call or write.
\item There is no one I can turn to.
\item I am no longer close to anyone.
\item My interest and ideas are not shared by those around me.
\item I feel left out.
\item I feel completely alone.
\item I am unable to reach out and communicate with those around me.
\item My social relationships are superficial.
\item I feel starved for company.
\item No one really knows me well.
\item I feel isolated from others.
\item I am unhappy being so withdrawn.
\item It is difficult for me to make friends.
\item I feel shut out and excluded by others.
\item People are around me but not with me.
\end{enumerate}

\subsubsection{Social Health}
Please indicate how you feel about the following aspects of your life. [Rating scale: \textit{1 = Strongly disagree, 2 = Moderately disagree, 3 = Slightly disagree, 4 = Neutral, 5 = Slightly agree, 6 = Moderately agree, 7 = Strongly agree}]

\begin{enumerate}
\item I do things that are meaningful to me.
\item I am able to take care of my needs.
\item I am able to handle things when they go wrong.
\item I am able to do things that I want to do.
\item I am happy with the friendships that I have.
\item I have people with whom I can do enjoyable things.
\item I feel I belong in my community.
\item In a crisis, I would have the support I need from family or friends.
\end{enumerate}

How do you feel about... [Rating scale: \textit{1 = Very displeased, 2 = Moderately displeased, 3 = Slightly displeased, 4 = Neutral, 5 = Slightly pleased, 6 = Moderately pleased, 7 = Very pleased}]

\begin{enumerate}
\item ...the way you spend your spare time.
\item ...the things you do with other people.
\item ...the people you see socially.
\item ...the quality of friendship in your life.
\item ...your personal understanding of yourself.
\item ...your creative or personal expression.
\item ...your relationships with parents, siblings, and other relatives.
\item ...your relationship with your spouse or significant other.
\item ...your relationships with friends.
\end{enumerate}

How would you rate your overall social health? [Rating scale: \textit{1 = Very unhealthy, 2 = Moderately unhealthy, 3 = Slightly unhealthy, 4 = Neither unhealthy nor healthy, 5 = Slightly healthy, 6 = Moderately healthy, 7 = Very healthy}]

\begin{enumerate}
\item My social health is...
\end{enumerate}

\subsubsection{Anthropomorphism} This measure was composed of four scales, listed below.

\textbf{\textit{Experience}}
In my opinion, the [Replika chatbot has/word games have] the capacity to feel:  [Rating scale: \textit{1 = Strongly disagree, 2 = Moderately disagree, 3 = Slightly disagree, 4 = Neutral, 5 = Slightly agree, 6 = Moderately agree, 7 = Strongly agree}]

\begin{enumerate}
\item Pain.
\item Fear.
\item Hunger.
\item Love.
\item Anger.
\item Pleasure.
\end{enumerate}

\textbf{\textit{Consciousness}}
In my opinion, the [Replika chatbot has/word games have]:  [Rating scale: \textit{1 = Strongly disagree, 2 = Moderately disagree, 3 = Slightly disagree, 4 = Neutral, 5 = Slightly agree, 6 = Moderately agree, 7 = Strongly agree}]
\begin{enumerate}
\item Consciousness of [itself/themselves].
\item Consciousness of me.
\item An understanding that I am conscious.
\item Consciousness of the world around [it/them].
\end{enumerate}

\textbf{\textit{Agency}}
In my opinion, the [Replika chatbot has/word games have] the capacity to:  [Rating scale: \textit{1 = Strongly disagree, 2 = Moderately disagree, 3 = Slightly disagree, 4 = Neutral, 5 = Slightly agree, 6 = Moderately agree, 7 = Strongly agree}]

\begin{enumerate}
\item Plan actions.
\item Exercise self-control.
\item Remember.
\item Act morally.
\item Act immorally.
\end{enumerate}

\textbf{\textit{Human Likeness}}
Please drag the slider to indicate how you view the [Replika chatbot/word games].

\begin{enumerate}
\item I view the [Replika chatbot/word games] as: [Sliding scale: \textit{1 = Fake, 7 = Natural}]
\item I view the [Replika chatbot/word games] as: [Sliding scale: \textit{1 = Not conscious, 7 = Conscious}]
\item I view the [Replika chatbot/word games] as: [Sliding scale: \textit{1 = Machinelike, 7 = Humanlike}]
\item I view the [Replika chatbot/word games] as: [Sliding scale: \textit{1 = Artificial, 7 = Lifelike}]
\item I view the [Replika chatbot/word games] as: [Sliding scale: \textit{1 = Dead, 7 = Alive}]
\item I view the [Replika chatbot/word games] as: [Sliding scale: \textit{1 = Stagnant, 7 = Lively}]
\item I view the [Replika chatbot/word games] as: [Sliding scale: \textit{1 = Apathetic, 7 = Responsive}]
\end{enumerate}

\subsubsection{Social Impact}
Please rate how harmful or helpful your engagement with the [Replika chatbot/word games] has been for your: [Rating scale: \textit{1 = Very harmful, 2 = Moderately harmful, 3 = Slightly harmful, 4 = Neutral, 5 = Slightly helpful, 6 = Moderately helpful, 7 = Very helpful}]

\begin{enumerate}
\item Social interactions.
\item Relationships with family or friends.
\end{enumerate}

\section{Acknowledgments}
R.G. and M.G. designed the study. R.G. collected and analyzed the data. R.G. and M.G. wrote the paper. The authors have no competing interests to declare.

This research was generously supported by the Data-Driven Social Science Initiative at Princeton University under Grant No. 24400-A0006-FA010 and the Program in Cognitive Science at Princeton University under Grant No. 28500-A0001-GR015. We thank these contributors for making this intensive longitudinal study possible.

This material is based upon work supported by the National Science Foundation Graduate Research Fellowships Program (GRFP) under Grant No. KB0013612. Any opinions, findings, and conclusions or recommendations expressed in this material are those of the authors and do not necessarily reflect the views of the National Science Foundation.

\bigskip

\bibliography{longitudinal-bib}

\begin{thebibliography}{56}
\providecommand{\natexlab}[1]{#1}

\bibitem[{{Anderson-Butcher}, Iachini, and Amorose(2008)}]{anderson-butcherInitialReliabilityValidity2008a}
{Anderson-Butcher}, D.; Iachini; and Amorose, A.~J. 2008.
\newblock Initial {{Reliability}} and {{Validity}} of the {{Perceived Social Competence Scale}}.
\newblock \emph{Research on Social Work Practice}, 18(1).

\bibitem[{Balch(2020)}]{balchAIMeFriendship2020}
Balch, O. 2020.
\newblock {{AI}} and Me: Friendship Chatbots Are on the Rise, but Is There a Gendered Design Flaw?
\newblock \emph{The Guardian}.

\bibitem[{Bartneck et~al.(2009)Bartneck, Kuli{\'c}, Croft, and Zoghbi}]{bartneckMeasurementInstrumentsAnthropomorphism2009}
Bartneck, C.; Kuli{\'c}, D.; Croft, E.; and Zoghbi, S. 2009.
\newblock Measurement {{Instruments}} for the {{Anthropomorphism}}, {{Animacy}}, {{Likeability}}, {{Perceived Intelligence}}, and {{Perceived Safety}} of {{Robots}}.
\newblock \emph{International Journal of Social Robotics}, 1(1): 71--81.

\bibitem[{Bartz, Tchalova, and Fenerci(2016)}]{bartzRemindersSocialConnection2016}
Bartz, J.~A.; Tchalova, K.; and Fenerci, C. 2016.
\newblock Reminders of Social Connection Can Attenuate Anthropomorphism: {{A}} Replication and Extension of {{Epley}}, {{Akalis}}, {{Waytz}}, and {{Cacioppo}} (2008).
\newblock \emph{Psychological Science}, 27(12): 1644--1650.

\bibitem[{Baumeister and Leary(1995)}]{baumeisterNeedBelongDesire1995}
Baumeister, R.~F.; and Leary, M.~R. 1995.
\newblock The Need to Belong: {{Desire}} for Interpersonal Attachments as a Fundamental Human Motivation.
\newblock \emph{Psychological Bulletin}, 117(3): 497--529.

\bibitem[{Br{\"u}ne and {Br{\"u}ne-Cohrs}(2006)}]{bruneTheoryMindEvolution2006}
Br{\"u}ne, M.; and {Br{\"u}ne-Cohrs}, U. 2006.
\newblock Theory of Mind---Evolution, Ontogeny, Brain Mechanisms and Psychopathology.
\newblock \emph{Neuroscience \& Biobehavioral Reviews}, 30(4): 437--455.

\bibitem[{Cacioppo and Cacioppo(2014)}]{cacioppoSocialRelationshipsHealth2014}
Cacioppo, J.~T.; and Cacioppo, S. 2014.
\newblock Social {{Relationships}} and {{Health}}: {{The Toxic Effects}} of {{Perceived Social Isolation}}.
\newblock \emph{Social and Personality Psychology Compass}, 8(2): 58--72.

\bibitem[{Cacioppo and Cacioppo(2018)}]{cacioppoGrowingProblemLoneliness2018}
Cacioppo, J.~T.; and Cacioppo, S. 2018.
\newblock The Growing Problem of Loneliness.
\newblock \emph{Lancet (London, England)}, 391(10119): 426.

\bibitem[{Cacioppo and Patrick(2008)}]{cacioppoLonelinessHumanNature2008}
Cacioppo, J.~T.; and Patrick, W. 2008.
\newblock \emph{Loneliness: {{Human}} Nature and the Need for Social Connection}.
\newblock WW Norton \& Company.

\bibitem[{Carlson et~al.(2011)Carlson, Sarkin, Levack, Sklar, Tally, Gilmer, and Groessl}]{carlsonEvaluatingMeasureSocial2011}
Carlson, J.~A.; Sarkin, A.~J.; Levack, A.~E.; Sklar, M.; Tally, S.~R.; Gilmer, T.~P.; and Groessl, E.~J. 2011.
\newblock Evaluating a {{Measure}} of {{Social Health Derived}} from {{Two Mental Health Recovery Measures}}: {{The California Quality}} of {{Life}} ({{CA-QOL}}) and {{Mental Health Statistics Improvement Program Consumer Survey}} ({{MHSIP}}).
\newblock \emph{Community Mental Health Journal}, 47(4): 454--462.

\bibitem[{Chandra et~al.(2025)Chandra, Hernandez, Ramos, Ershadi, Bhattacharjee, Amores, Okoli, Paradiso, Warreth, and Suh}]{chandraLongitudinalStudySocial2025}
Chandra, M.; Hernandez, J.; Ramos, G.; Ershadi, M.; Bhattacharjee, A.; Amores, J.; Okoli, E.; Paradiso, A.; Warreth, S.; and Suh, J. 2025.
\newblock Longitudinal {{Study}} on {{Social}} and {{Emotional Use}} of {{AI Conversational Agent}}.
\newblock arXiv:2504.14112.

\bibitem[{Colombatto and Fleming(2023)}]{colombattoFolkPsychologicalAttributions2023}
Colombatto, C.; and Fleming, S.~M. 2023.
\newblock Folk {{Psychological Attributions}} of {{Consciousness}} to {{Large Language Models}}.
\newblock Preprint, PsyArXiv.

\bibitem[{Cost(2023)}]{costMarriedFatherCommits2023}
Cost, B. 2023.
\newblock Married Father Commits Suicide after Encouragement by {{AI}} Chatbot: Widow.
\newblock \emph{New York Post}.

\bibitem[{Croes and Antheunis(2021)}]{croesCanWeBe2021}
Croes, E. A.~J.; and Antheunis, M.~L. 2021.
\newblock Can We Be Friends with {{Mitsuku}}? {{A}} Longitudinal Study on the Process of Relationship Formation between Humans and a Social Chatbot.
\newblock \emph{Journal of Social and Personal Relationships}, 38(1): 279--300.

\bibitem[{Epley et~al.(2008{\natexlab{a}})Epley, Akalis, Waytz, and Cacioppo}]{epleyCreatingSocialConnection2008}
Epley, N.; Akalis, S.; Waytz, A.; and Cacioppo, J.~T. 2008{\natexlab{a}}.
\newblock Creating Social Connection through Inferential Reproduction: {{Loneliness}} and Perceived Agency in Gadgets, Gods, and Greyhounds.
\newblock \emph{Psychological Science}, 19(2): 114--120.

\bibitem[{Epley et~al.(2008{\natexlab{b}})Epley, Waytz, Akalis, and Cacioppo}]{epleyWhenWeNeed2008}
Epley, N.; Waytz, A.; Akalis, S.; and Cacioppo, J. 2008{\natexlab{b}}.
\newblock When {{We Need A Human}}: {{Motivational Determinants}} of {{Anthropomorphism}}.
\newblock \emph{Social Cognition - SOC COGNITION}, 26: 143--155.

\bibitem[{Epley, Waytz, and Cacioppo(2007)}]{epleySeeingHumanThreefactor2007}
Epley, N.; Waytz, A.; and Cacioppo, J.~T. 2007.
\newblock On Seeing Human: {{A}} Three-Factor Theory of Anthropomorphism.
\newblock \emph{Psychological Review}, 114(4): 864--886.

\bibitem[{Eyssel and Reich(2013)}]{eysselLonelinessMakesHeart2013}
Eyssel, F.; and Reich, N. 2013.
\newblock Loneliness Makes the Heart Grow Fonder (of Robots): On the Effects of Loneliness on Psychological Anthropomorphism.
\newblock In \emph{Proceedings of the 8th {{ACM}}/{{IEEE}} International Conference on {{Human-robot}} Interaction}, {{HRI}} '13, 121--122. Tokyo, Japan: IEEE Press.
\newblock ISBN 978-1-4673-3055-8.

\bibitem[{Eyssel and Pfundmair(2015)}]{eysselPredictorsPsychologicalAnthropomorphization2015}
Eyssel, F.~A.; and Pfundmair, M. 2015.
\newblock Predictors of Psychological Anthropomorphization, Mind Perception, and the Fulfillment of Social Needs: {{A}} Case Study with a Zoomorphic Robot.
\newblock \emph{2015 24th IEEE International Symposium on Robot and Human Interactive Communication (RO-MAN)}, 827--832.

\bibitem[{Fang et~al.(2025)Fang, Liu, Danry, Lee, Chan, Pataranutaporn, Maes, Phang, Lampe, Ahmad, and Agarwal}]{fangHowAIHuman2025}
Fang, C.~M.; Liu, A.~R.; Danry, V.; Lee, E.; Chan, S. W.~T.; Pataranutaporn, P.; Maes, P.; Phang, J.; Lampe, M.; Ahmad, L.; and Agarwal, S. 2025.
\newblock How {{AI}} and {{Human Behaviors Shape Psychosocial Effects}} of {{Chatbot Use}}: {{A Longitudinal Randomized Controlled Study}}.
\newblock arXiv:2503.17473.

\bibitem[{Franke, Attig, and Wessel(2019)}]{frankePersonalResourceTechnology2019}
Franke, T.; Attig, C.; and Wessel, D. 2019.
\newblock A Personal Resource for Technology Interaction: {{Development}} and Validation of the {{Affinity}} for {{Technology Interaction}} ({{ATI}}) Scale.
\newblock \emph{International Journal of Human-Computer Interaction}, 35(6): 456--467.

\bibitem[{Frith(2002)}]{frithAttentionActionAwareness2002}
Frith, C. 2002.
\newblock Attention to Action and Awareness of Other Minds.
\newblock \emph{Consciousness and Cognition: An International Journal}, 11(4): 481--487.

\bibitem[{Gosling, Rentfrow, and Swann~Jr.(2003)}]{goslingVeryBriefMeasure2003}
Gosling, S.~D.; Rentfrow, P.~J.; and Swann~Jr., W.~B. 2003.
\newblock A Very Brief Measure of the {{Big-Five}} Personality Domains.
\newblock \emph{Journal of Research in Personality}, 37(6): 504--528.

\bibitem[{Gray, Gray, and Wegner(2007)}]{grayDimensionsMindPerception2007}
Gray, H.~M.; Gray, K.; and Wegner, D.~M. 2007.
\newblock Dimensions of {{Mind Perception}}.
\newblock \emph{Science}, 315(5812): 619--619.

\bibitem[{Graziano(2013)}]{grazianoConsciousnessSocialBrain2013}
Graziano, M. S.~A. 2013.
\newblock \emph{Consciousness and the Social Brain}.
\newblock Consciousness and the Social Brain. New York, NY, US: Oxford University Press.
\newblock ISBN 978-0-19-992864-4.

\bibitem[{Guingrich and Graziano(2024)}]{guingrichAscribingConsciousnessArtificial2024}
Guingrich, R.~E.; and Graziano, M. S.~A. 2024.
\newblock Ascribing Consciousness to Artificial Intelligence: Human-{{AI}} Interaction and Its Carry-over Effects on Human-Human Interaction.
\newblock \emph{Frontiers in Psychology}, 15.

\bibitem[{Guingrich and Graziano(2025{\natexlab{a}})}]{guingrichChatbotsSocialCompanions2025}
Guingrich, R.~E.; and Graziano, M. S.~A. 2025{\natexlab{a}}.
\newblock Chatbots as {{Social Companions}}: {{How People Perceive Consciousness}}, {{Human Likeness}}, and {{Social Health Benefits}} in {{Machines}}.
\newblock In Hacker, P., ed., \emph{Oxford {{Intersections}}: {{AI}} in {{Society}}}. Oxford University Press.
\newblock ISBN 978-0-19-894521-5.

\bibitem[{Guingrich and Graziano(2025{\natexlab{b}})}]{guingrichPdoomAIOptimism2025}
Guingrich, R.~E.; and Graziano, M. S.~A. 2025{\natexlab{b}}.
\newblock P(Doom) {{Versus AI Optimism}}: {{Attitudes Toward Artificial Intelligence}} and the {{Factors That Shape Them}}.
\newblock \emph{Journal of Technology in Behavioral Science}.

\bibitem[{Kim and McGill(2023)}]{kimAIinducedDehumanization2023}
Kim, H.-y.; and McGill, A.~L. 2023.
\newblock {{AI-induced}} Dehumanization.
\newblock \emph{Journal of Consumer Psychology}, n/a(n/a).

\bibitem[{Kothe and Ling(2019)}]{kotheRetentionParticipantsRecruited2019}
Kothe, E.~J.; and Ling, M. 2019.
\newblock Retention of Participants Recruited to a Multi-Year Longitudinal Study via {{Prolific}}.
\newblock Preprint, PsyArXiv.

\bibitem[{Koverola et~al.(2022)Koverola, Kunnari, Sundvall, and Laakasuo}]{koverolaGeneralAttitudesRobots2022}
Koverola, M.; Kunnari, A.; Sundvall, J.; and Laakasuo, M. 2022.
\newblock General {{Attitudes Towards Robots Scale}} ({{GAToRS}}): {{A New Instrument}} for {{Social Surveys}}.
\newblock \emph{International Journal of Social Robotics}, 14(7): 1559--1581.

\bibitem[{Lieberz et~al.(2021)Lieberz, Shamay-Tsoory, Saporta, Esser, Kuskova, Stoffel-Wagner, Hurlemann, and Scheele}]{lieberzLonelinessSocialBrain2021}
Lieberz, J.; Shamay-Tsoory, S.~G.; Saporta, N.; Esser, T.; Kuskova, E.; Stoffel-Wagner, B.; Hurlemann, R.; and Scheele, D. 2021.
\newblock Loneliness and the {{Social Brain}}: {{How Perceived Social Isolation Impairs Human Interactions}}.
\newblock \emph{Advanced Science}, 8(21): 2102076.

\bibitem[{Malfacini(2025)}]{malfaciniImpactsCompanionAI2025}
Malfacini, K. 2025.
\newblock The Impacts of Companion {{AI}} on Human~Relationships: Risks, Benefits, and Design Considerations.
\newblock \emph{AI \& SOCIETY}.

\bibitem[{Maples et~al.(2024)Maples, Cerit, Vishwanath, and Pea}]{maplesLonelinessSuicideMitigation2024}
Maples, B.; Cerit, M.; Vishwanath, A.; and Pea, R. 2024.
\newblock Loneliness and Suicide Mitigation for Students Using {{GPT3-enabled}} Chatbots.
\newblock \emph{npj Mental Health Research}, 3(1): 1--6.

\bibitem[{Messer and Harter(2012)}]{messerSelfperceptionProfileAdults2012}
Messer, B.; and Harter, S. 2012.
\newblock The Self-Perception Profile for Adults: {{Manual}} and Questionnaires.

\bibitem[{Nagel(1974)}]{nagelWhatItBe1974}
Nagel, T. 1974.
\newblock What {{Is It Like}} to {{Be}} a {{Bat}}?
\newblock \emph{The Philosophical Review}, 83(4): 435--450.

\bibitem[{Nass, Steuer, and Siminoff(1994)}]{nassComputersAreSocial1994}
Nass, C.; Steuer, J.; and Siminoff, E. 1994.
\newblock Computers Are {{Social Actors}}.
\newblock In \emph{Conference on {{Human Factors}} in {{Computing Systems}} - {{Proceedings}}}, 204.

\bibitem[{{Office of the U.S. Surgeon General}(2023)}]{officeoftheu.s.surgeongeneralOurEpidemicLoneliness2023}
{Office of the U.S. Surgeon General}. 2023.
\newblock Our {{Epidemic}} of {{Loneliness}} and {{Isolation}}: {{The U}}.{{S}}. {{Surgeon General}}'s {{Advisory}} on the {{Healing Effects}} of {{Social Connection}} and {{Community}}.
\newblock Technical report, United States Public Health Service.

\bibitem[{O'Sullivan et~al.(2021)O'Sullivan, Burns, Leavey, Leroi, Burholt, Lubben, {Holt-Lunstad}, Victor, Lawlor, {Vilar-Compte}, Perissinotto, Tully, Sullivan, Rosato, Power, Tiilikainen, and Prohaska}]{osullivanImpactCOVID19Pandemic2021}
O'Sullivan, R.; Burns, A.; Leavey, G.; Leroi, I.; Burholt, V.; Lubben, J.; {Holt-Lunstad}, J.; Victor, C.; Lawlor, B.; {Vilar-Compte}, M.; Perissinotto, C.~M.; Tully, M.~A.; Sullivan, M.~P.; Rosato, M.; Power, J.~M.; Tiilikainen, E.; and Prohaska, T.~R. 2021.
\newblock Impact of the {{COVID-19 Pandemic}} on {{Loneliness}} and {{Social Isolation}}: {{A Multi-Country Study}}.
\newblock \emph{International Journal of Environmental Research and Public Health}, 18(19): 9982.

\bibitem[{Premack and Woodruff(1978)}]{premackDoesChimpanzeeHave1978}
Premack, D.; and Woodruff, G. 1978.
\newblock Does the Chimpanzee Have a Theory of Mind?
\newblock \emph{Behavioral and Brain Sciences}, 1(4): 515--526.

\bibitem[{Prinz(2017)}]{prinzModelingSelfOthers2017}
Prinz, W. 2017.
\newblock Modeling Self on Others: {{An}} Import Theory of Subjectivity and Selfhood.
\newblock \emph{Consciousness and Cognition: An International Journal}, 49: 347--362.

\bibitem[{Reeves and Nass(1996)}]{reevesMediaEquationHow1996}
Reeves, B.; and Nass, C. 1996.
\newblock The {{Media Equation}}: {{How People Treat Computers}}, {{Television}}, and {{New Media Like Real People}} and {{Pla}}.
\newblock \emph{Bibliovault OAI Repository, the University of Chicago Press}.

\bibitem[{Roose(2024)}]{rooseCanAIBe2024}
Roose, K. 2024.
\newblock Can {{A}}.{{I}}. {{Be Blamed}} for a {{Teen}}'s {{Suicide}}?
\newblock \emph{The New York Times}.

\bibitem[{Rosseel(2012)}]{rosseelLavaanPackageStructural2012}
Rosseel, Y. 2012.
\newblock Lavaan: {{An R Package}} for {{Structural Equation Modeling}}.
\newblock \emph{Journal of Statistical Software}, 48: 1--36.

\bibitem[{Russell, Peplau, and Cutrona(1980)}]{russellRevisedUCLALoneliness1980}
Russell, D.; Peplau, L.~A.; and Cutrona, C.~E. 1980.
\newblock The Revised {{UCLA Loneliness Scale}}: {{Concurrent}} and Discriminant Validity Evidence.
\newblock \emph{Journal of Personality and Social Psychology}, 39(3): 472--480.

\bibitem[{Scott et~al.(2023)Scott, Neumann, Niess, and Wo{\'z}niak}]{scottYouMindUser2023}
Scott, A.~E.; Neumann, D.; Niess, J.; and Wo{\'z}niak, P.~W. 2023.
\newblock Do {{You Mind}}? {{User Perceptions}} of {{Machine Consciousness}}.
\newblock In \emph{Proceedings of the 2023 {{CHI Conference}} on {{Human Factors}} in {{Computing Systems}}}, {{CHI}} '23, 1--19. New York, NY, USA: Association for Computing Machinery.
\newblock ISBN 978-1-4503-9421-5.

\bibitem[{Seyfarth and Cheney(2013)}]{seyfarthAffiliationEmpathyOrigins2013}
Seyfarth, R.~M.; and Cheney, D.~L. 2013.
\newblock Affiliation, Empathy, and the Origins of Theory of Mind.
\newblock \emph{Proceedings of the National Academy of Sciences of the United States of America}, 110(Suppl 2): 10349--10356.

\bibitem[{Skjuve, F{\o}lstad, and Brandtz{\ae}g(2023)}]{skjuveLongitudinalStudySelfDisclosure2023}
Skjuve, M.; F{\o}lstad, A.; and Brandtz{\ae}g, P.~B. 2023.
\newblock A {{Longitudinal Study}} of {{Self-Disclosure}} in {{Human}}--{{Chatbot Relationships}}.
\newblock \emph{Interacting with Computers}, 35(1): 24--39.

\bibitem[{Skjuve et~al.(2022)Skjuve, F{\o}lstad, Fostervold, and Brandtzaeg}]{skjuveLongitudinalStudyHuman2022}
Skjuve, M.; F{\o}lstad, A.; Fostervold, K.~I.; and Brandtzaeg, P.~B. 2022.
\newblock A Longitudinal Study of Human--Chatbot Relationships.
\newblock \emph{International Journal of Human-Computer Studies}, 168: 102903.

\bibitem[{Tomova, Andrews, and Blakemore(2021)}]{tomovaImportanceBelongingAvoidance2021}
Tomova, L.; Andrews, J.~L.; and Blakemore, S.-J. 2021.
\newblock The Importance of Belonging and the Avoidance of Social Risk Taking in Adolescence.
\newblock \emph{Developmental Review}, 61: 100981.

\bibitem[{Tsoi et~al.(2021)Tsoi, Hamlin, Waytz, Baron, and Young}]{tsoiCooperationAdvantageTheory2021}
Tsoi, L.; Hamlin, J.~K.; Waytz, A.; Baron, A.~S.; and Young, L.~L. 2021.
\newblock A {{Cooperation Advantage}} for {{Theory}} of {{Mind}} in {{Children}} and {{Adults}}.
\newblock https://guilfordjournals.com/doi/10.1521/soco.2021.39.1.19.

\bibitem[{Turkle(2007)}]{turkleAuthenticityAgeDigital2007}
Turkle, S. 2007.
\newblock Authenticity in the Age of Digital Companions.
\newblock \emph{Interaction Studies}, 8(3): 501--517.

\bibitem[{Williams(2009)}]{williamsChapter6Ostracism2009}
Williams, K.~D. 2009.
\newblock Chapter 6 {{Ostracism}}: {{A Temporal Need}}-{{Threat Model}}.
\newblock In \emph{Advances in {{Experimental Social Psychology}}}, volume~41, 275--314. Academic Press.

\bibitem[{Wimmer and Perner(1983)}]{wimmerBeliefsBeliefsRepresentation1983}
Wimmer, H.; and Perner, J. 1983.
\newblock Beliefs about Beliefs: {{Representation}} and Constraining Function of Wrong Beliefs in Young Children's Understanding of Deception.
\newblock \emph{Cognition}, 13(1): 103--128.

\bibitem[{Xie and Pentina(2022)}]{xieAttachmentTheoryFramework2022}
Xie, T.; and Pentina, I. 2022.
\newblock Attachment {{Theory}} as a {{Framework}} to {{Understand Relationships}} with {{Social Chatbots}}: {{A Case Study}} of {{Replika}}.
\newblock In \emph{Proceedings of the 55th {{Hawaii International Conference}} on {{System Sciences}}}, {{HICSS}} '55, 2046--2055. Hawaii, USA: Hawaii International Conference on System Sciences.
\newblock ISBN 978-0-9981331-5-7.

\bibitem[{Young and Monroe(2019)}]{youngAutonomousMoralsInferences2019}
Young, A.; and Monroe, A. 2019.
\newblock Autonomous Morals: {{Inferences}} of Mind Predict Acceptance of {{AI}} Behavior in Sacrificial Moral Dilemmas.
\newblock \emph{Journal of Experimental Social Psychology}, 85: 103870.

\end{thebibliography}

\end{document}